\documentclass{article}
\usepackage{cite}
\usepackage{graphicx}
\usepackage{dcolumn}

\input{tcilatex}
\begin{document}

\date{}
\title{About the effects of rotation on the Landau levels in an elastic medium
with a spiral dislocation}
\author{Paolo Amore\thanks{%
paolo@ucol.mx} \\
Facultad de Ciencias, Universidad de Colima, \\
Bernal D\'{\i}az del Castillo 340, Colima, Colima, M\'exico \\
\\
Francisco M. Fern\'andez \thanks{%
fernande@quimica.unlp.edu.ar} \\
INIFTA, Diag. 113 y 64 S/N, \\
Sucursal 4, Casilla de Correo 16, 1900 La Plata, Argentina}
\maketitle

\begin{abstract}
In this Paper we analyze a model proposed recently with the
purpose of studying the effects of rotation on the interaction of
a point charge with a uniform magnetic field in an elastic medium
with a spiral dislocation. In particular we focus on the
approximation proposed by the authors that consists of changing
the left boundary condition in order to obtain analytical results.
We show that this approximation leads to quantitative and
qualitative errors, the most relevant one being a wrong prediction
of the level spacing.
\end{abstract}

\section{Introduction}

\label{sec:intro}

In a recent paper Maia and Bakke\cite{MB20} analyzed the effects of rotation
on the interaction of a point charge with a uniform magnetic field in an
elastic medium with a spiral dislocation. In order to obtain analytical
solutions to the Schr\"{o}dinger equation in a rotating frame with a
constant angular velocity Maia and Bakke\cite{MB20} changed the boundary
conditions of the model. Based on this approximation they concluded that
both the topology of the defect and rotation modify the degeneracy of the
Landau levels.

The purpose of this paper is to analyze the effect of the change of the left
boundary condition on the results because Maia and Bakke\cite{MB20} did not
discuss this point in detail and their weak argument leaves much to be
desired. In section~\ref{sec:analitical} we derive the analytical results by
means of the Frobenius method because it is clearer than the approach based
on the confluent hypergeometric function followed by Maia and Bakke\cite
{MB20}. In section~\ref{sec:exact_BC} we solve the eigenvalue equation with
the correct left boundary condition and compare the numerical results thus
obtained with the analytical ones. Finally, in section~\ref{sec:conclusions}
we summarize the main results and draw conclusions.

\section{Analytical results}

\label{sec:analitical}

In this section we outline some of the results derived by Maia and Bakke\cite
{MB20}. Our starting point is the radial differential equation
\begin{eqnarray}
&&\left( 1+\frac{\beta ^{2}}{r^{2}}\right) h^{\prime \prime }+\left( \frac{1%
}{r}-\frac{\beta ^{2}}{r^{3}}\right) h^{\prime }-\frac{l^{2}+m\omega l\beta
^{2}}{r^{2}+\beta ^{2}}h-\frac{m^{2}\delta ^{2}r^{4}}{4\left( r^{2}+\beta
^{2}\right) }h  \nonumber \\
&&-\frac{m^{2}\beta ^{2}\left( \delta ^{2}-\omega ^{2}\right) r^{2}}{4\left(
r^{2}+\beta ^{2}\right) }h+\left[ 2m\left( E+\Omega l\right) -k^{2}+m\omega
l\right] h=0,  \label{eq:dif_eq_h}
\end{eqnarray}
derived by those authors. By means of the change of variables $y=m\delta
\left( r^{2}+\beta ^{2}\right) /2$ they obtained the simpler equation
\begin{eqnarray}
&&y^{2}g^{\prime \prime }+yg^{\prime }-\frac{\gamma ^{2}}{4}g-\frac{y^{2}}{4}%
g+\tau yg=0,  \nonumber \\
&&\gamma =l+\frac{m\omega \beta ^{2}}{2},  \nonumber \\
&&\tau =\frac{1}{2m\delta }\left[ 2m\left( E+\Omega l\right) -k^{2}+m\omega
l+\frac{m^{2}\left( \delta ^{2}+\omega ^{2}\right) \beta ^{2}}{4}\right] .
\label{eq:dif_eq_g}
\end{eqnarray}
Maia and Bakke\cite{MB20} resorted to the same function $h$ for equations (%
\ref{eq:dif_eq_h}) and (\ref{eq:dif_eq_g}). However, we decided to use a
different name because $h(z)$ and $g(z)$ are obviously different because $%
h(r)=\left( g\circ y\right) (r)=g(y(r))$. The authors stated that
``Henceforth, let us impose that $h(y)\rightarrow 0$ when $y\rightarrow
\infty $ and $y\rightarrow 0$. Note that, since $0<\beta <1$, then, we can
assume that $\beta ^{2}<<1$. Thus, when $r\rightarrow 0$, we can consider $%
y\rightarrow 0$ without loss of generality [13].'' In what follows we
estimate the effect of such drastic change of the left boundary condition.

Maia and Bakke\cite{MB20} derived the allowed values of $\tau $ by writing $%
g(y)$ in terms of the confluent hypergeometric function. Here, we resort to
the Frobenius method and write
\begin{equation}
g(y)=y^{|\gamma |/2}e^{-y/2}\sum_{j=0}^{\infty }c_{j}y^{j},
\label{eq:g(y)_Frobenius}
\end{equation}
that leads to the following recurrence relation for the coefficients:
\begin{equation}
c_{j+1}=A_{j}c_{j},\;A_{j}=\frac{\left| \gamma \right| +2j-2\tau +1}{2\left(
j+1\right) \left( \left| \gamma \right| +j+1\right) }.  \label{eq:rec_rel}
\end{equation}
In order to obtain solutions with the correct behaviour when $y\rightarrow
\infty $ we have to choose $\tau $ so that the infinite series in equation (%
\ref{eq:g(y)_Frobenius}) terminates\cite{F21}. The requirement $c_{n}\neq 0$
and $c_{n+1}=0$, $n=0,1,\ldots $, leads to $c_{j}=c_{jn}=0$ for all $j>n$
and the exact solutions
\begin{equation}
g_{n}(y)=y^{|\gamma |/2}e^{-y/2}\sum_{j=0}^{n}c_{jn}y^{j},
\label{eq:g_jn(y)}
\end{equation}
for
\begin{equation}
\tau =\tau _{n}=\frac{1}{2}\left( 2n+1+|\gamma |\right) ,  \label{eq:tau_n}
\end{equation}
that agrees with the one derived by Maia and Bakke\cite{MB20}. The factor $%
A_{j}$ takes the simpler form
\begin{equation}
A_{j}=\frac{j-n}{\left( j+1\right) \left( \left| \gamma \right| +j+1\right) }%
.  \label{eq:A_(jn)}
\end{equation}

\section{Exact boundary conditions}

\label{sec:exact_BC}

When $r=0$ then $y=y_{0}=m\delta \beta ^{2}/2$ and the correct boundary
condition for the differential equation (\ref{eq:dif_eq_g}) is $g(y_{0})=0$.
The choice $y_{0}=0$ is rather unphysical because it leads to $r^{2}=-\beta
^{2}$; however, Maia and Bakke\cite{MB20} stated that they could resort to
this approximation \textit{without loss of generality} because $\beta
^{2}\ll 1$ as mentioned above. In what follows we analyze to which extent
this approximation is reasonable. To begin with, note that the behaviour at
origin of the solution to the differential equation (\ref{eq:dif_eq_h}) is $%
h\sim r^{2}$ while, on the other hand, the behaviour at origin of the
solution to (\ref{eq:dif_eq_g}) chosen by Maia and Bakke\cite{MB20} is $%
g\sim y^{|\gamma |/2}$. Under the approximation $\beta \sim 0$ it leads to $%
g\sim r^{|\gamma |}$ that is consistent with the correct asymptotic
behaviour at origin only when $|\gamma |=2$. Besides, $g^{\prime }(y)$
diverges at $y=0$ when $0<|\gamma |<2$.

In order to solve the differential equation (\ref{eq:dif_eq_g}) with the
proper boundary condition we resort to the shooting method. The choice of
suitable values for the model parameters is rather difficult because the
equations reported by Maia and Bakke\cite{MB20} do not exhibit unit
consistency. They did not indicate the chosen units explicitly but we assume
that they followed early papers in which they stated that ``we shall use the
units $\hbar =1$, $c=1$''\cite{MB19}. Unfortunately, this expression does
not mean anything because we do not know what are actually the units of
length, energy, etc,. (see \cite{F20} for a clear pedagogical discussion of
the subject and an earlier criticism of this undesirable practice\cite{F20b}%
). The fact is that all their equations lack of unit consistency. Consider,
for example, the expression for $\gamma $ in equation (\ref{eq:dif_eq_g})
where we appreciate that $l$ is an integer and $m\omega \beta ^{2}/2$
exhibits units of energy. It is not clear how the authors converted the
latter term into a dimensionless one because they did not explain it. As a
result the expression for the energy $E_{n,l,k}$ shown by Maia and Bakke\cite
{MB20} is extremely inconsistent. For example, the term $\omega l$ should be
multiplied by $\hbar $ in order to have units of energy; on the other hand $%
k^{2}/(2m)$ should be multiplied by $\hbar ^{2}$ in order to have the same
units. What should we do with the term $m\omega \Omega \beta ^{2}/2$ that
exhibits units of frequency$\times $energy?. Since it is not possible to
estimate physically reasonable values of $y_{0}$ and $\gamma $ we choose
them arbitrarily.

\begin{figure}[tbp]
\begin{center}
\includegraphics[width=9cm]{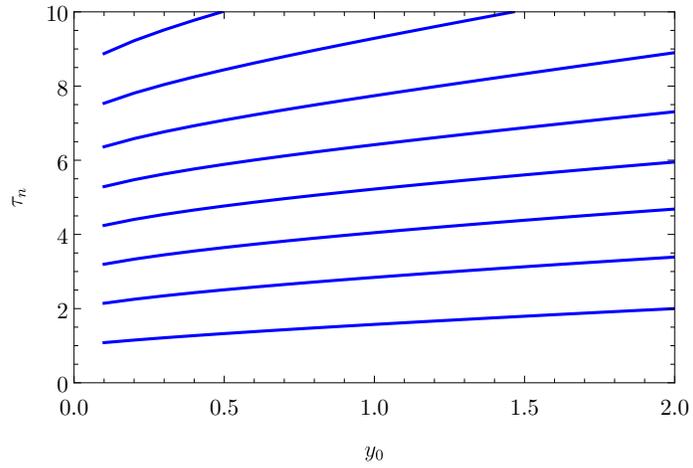}
\end{center}
\caption{Numerical eigenvalues $\tau_n$ vs $y_0$ for $\gamma=1$}
\label{fig:taun_y0}
\end{figure}

Figure~\ref{fig:taun_y0} shows that the eigenvalues $\tau _{n}$ increase
monotonously with $y_{0}$ and that this behaviour increases with the radial
quantum number. This effect has not been taken into account by Maia and Bakke%
\cite{MB20} because they only considered $y_{0}=0$.

\begin{figure}[tbp]
\begin{center}
\includegraphics[width=9cm]{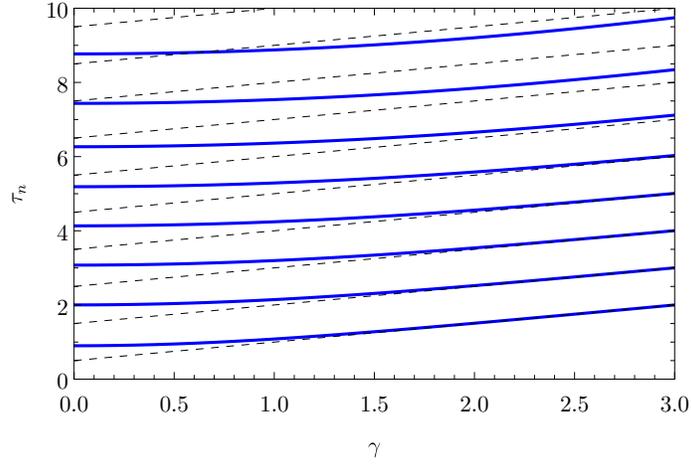}
\end{center}
\caption{Eigenvalues $\tau_n$ vs $\gamma$ for $y_0=0.1$. The continuous and
dashed lines indicate numerical and analytical ($y_0=0$) results,
respectively}
\label{fig:taun_gamma}
\end{figure}

Figure~\ref{fig:taun_gamma} shows that the eigenvalues $\tau _{n}$ increase
monotonously with $\gamma $, exactly as the analytical expressions (\ref
{eq:tau_n}). The discrepancy produced by the boundary condition at $y_{0}$
seems to be less relevant as $\gamma $ increases. We will discuss this point
in more detail below.

\begin{figure}[tbp]
\begin{center}
\includegraphics[width=9cm]{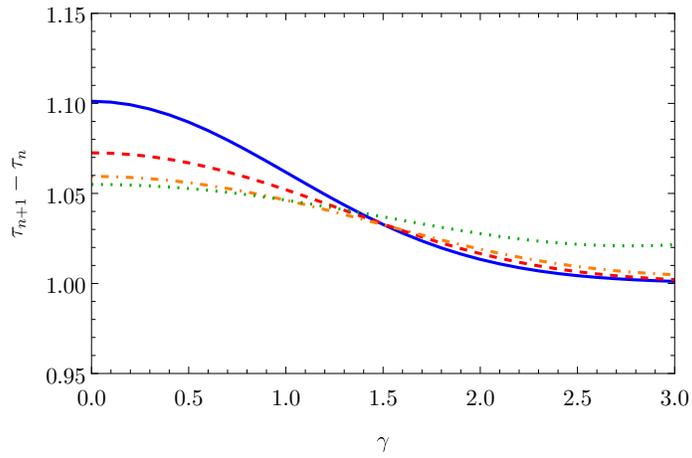}
\end{center}
\caption{$\tau_{n+1}-\tau_n$ vs $\gamma$ for $n=0$ (blue, continuous line), $%
n=1$ (red, dashed line), $n=2$ (orange, dash-point line) , $n=3$ (green,
point line) and $y_0=0.1$}
\label{fig:deltatau}
\end{figure}

The difference between two analytical eigenvalues (\ref{eq:tau_n}) is
independent of $\gamma $: $\tau _{n}-\tau _{j}=n-j$. Figure~\ref
{fig:deltatau} shows that if we choose the correct left boundary condition $%
\tau _{n+1}-\tau _{n}$, $n=0,1,2,3$, change with $\gamma $. The unphysical
approximation proposed by Maia and Bakke\cite{MB20} does not take into
account this fact. Note that the purpose of their paper was to determine the
effect of the dislocation given by the model parameter $\beta $ on the
Landau levels and they failed to do it properly. The variation of $\Delta
\tau $ with $\gamma $ may appear to be weak at first sight but it should be
taken into account that we have chosen a relatively small value of $%
y_{0}=0.1 $. We are unable to estimate reasonable physical values of $%
y_{0}=m\delta \beta ^{2}/2$ because Maia and Bakke\cite{MB20} did not show
suitable values for the model parameters and also because their equations
are inconsistent with respect to units.

\begin{figure}[tbp]
\begin{center}
\includegraphics[width=9cm]{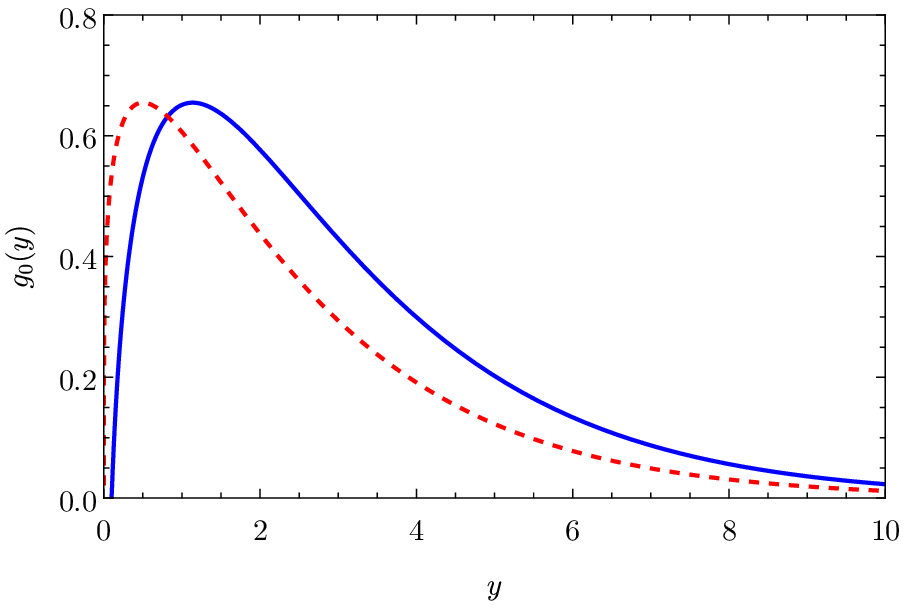}\\[0pt]
\includegraphics[width=9cm]{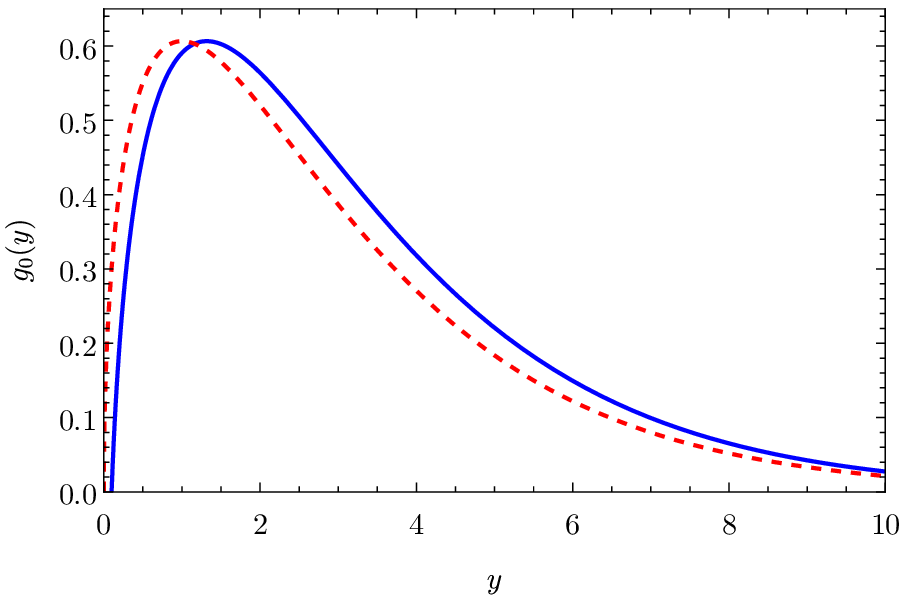}\\[0pt]
\includegraphics[width=9cm]{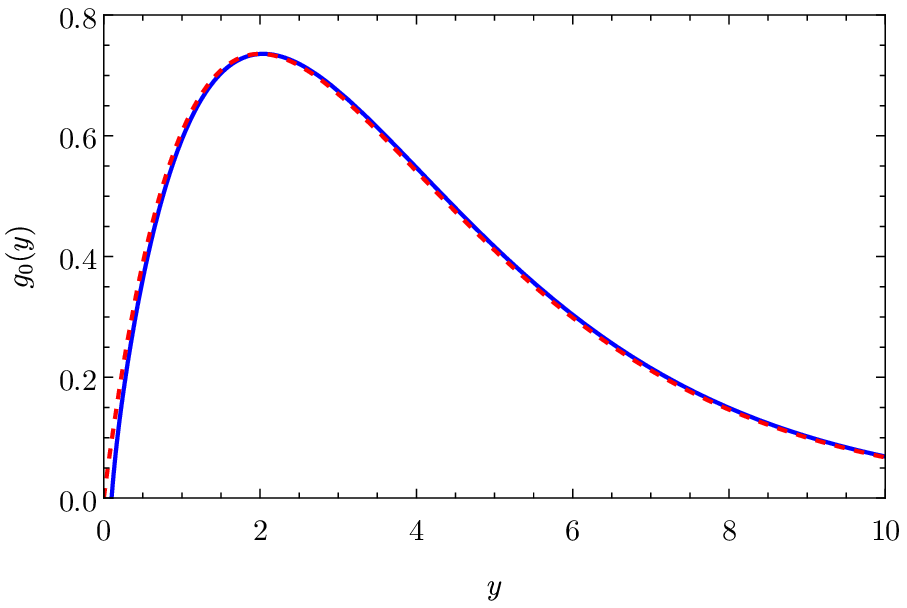}
\end{center}
\caption{Analytical (red dashed line, $y_0=0$) and numerical (blue
continuous line, $y_0=0.1$) values of $g_0(y)$ for $\gamma=0.5$, $\gamma=1$
and $\gamma=2$ (top to bottom)}
\label{Fig:g0}
\end{figure}

Figure \ref{Fig:g0} shows $g_{0}(y)$ for $\gamma =0.5$, $\gamma =1$ and $%
\gamma =2$ with the boundary conditions $g(0)=0$ and $g(0.1)=0$. We
appreciate that the agreement is better when $\gamma =2$ as argued above
(only in this case the maxima are reasonably close). As $\gamma $ increases
both kind of eigenfunctions become increasingly small in a neighbourhood of
the origin; consequently, if $y_{0}$ is small enough then the effect of the
Dirichlet boundary condition at $y_{0}$ is expected to be less noticeable.
Obviously, the discrepancy between both kinds of solutions increases as $%
y_{0}$ increases.

\section{Conclusions}

\label{sec:conclusions}

The purpose of this Paper is to analyze the effect of changing the
left boundary condition in the model proposed by Maia and
Bakke\cite{MB20}. Our results clearly show that the errors are not
only quantitative but also qualitative. If $y_{0}$ is small enough
the discrepancy between the solutions with Dirichlet boundary
conditions at $y=0$ and $y=y_{0}>0$ decreases as $\gamma $
decreases. The reason is that the larger the value of $\gamma $
the smaller the magnitude of the eigenfunctions in a neighbourhood
of the origin. In all our calculations we have chosen $y_{0}$
small because it is the condition under which the approach of Maia
and Bakke\cite{MB20} should work better. However, the main
drawback of their approximation is a wrong prediction of the level
spacing. Under the unphysical boundary condition at $y=0$ the
level spacing does not depend on $\gamma $ while the use of the
correct boundary condition at $y_{0}>0$ predicts that the level
spacing depends on that model parameter. Another relevant point is
that the equations derived by Maia and Bakke\cite{MB20} lack unit
consistency. For this reason they are completely useless for any
physical application. These authors carried out a similar mistake
in an earlier paper\cite{MB18} but in that case it was easier to
derive the equations properly\cite{F20b}.

\section*{Acknowledgements}

The research of P.A. was supported by Sistema nacional de Investigadores
(M\'exico)

\end{document}